\documentclass[lettersize,journal]{IEEEtran}
\usepackage{amsmath,amsfonts}
\usepackage{algorithmic}
\usepackage{algorithm}
\usepackage{array}
\usepackage[caption=false,font=normalsize,labelfont=sf,textfont=sf]{subfig}
\usepackage{textcomp}
\usepackage{stfloats}
\usepackage{url}
\usepackage{verbatim}
\usepackage{amssymb}
\usepackage{graphicx}
\def\BibTeX{{\rm B\kern-.05em{\sc i\kern-.025em b}\kern-.08em
    T\kern-.1667em\lower.7ex\hbox{E}\kern-.125emX}}
\usepackage{balance}
\begin{document}
\title{Blind Recognition of Polar Codes Using Successive Cancellation List Decoding}
\author{Changwei Tu, Yang Liu, Xianzhao Feng, and Kai Niu
\thanks{This work is supported by the National Natural Science Foundation of China under Grant 62321001 and Grant 62471054.
  \emph{(Corresponding author: Kai Niu.)}
  
The authors are with the Key Laboratory of Universal Wireless Communications, Ministry of Education,
Beijing University of Posts and Telecommunications, Beijing, China.
Email: \{tuchangwei, liuyang, foreseexz, niukai\}@bupt.edu.cn.

This work has been submitted to the IEEE for possible publication.
Copyright may be transferred without notice, after which this version may no longer be accessible.

}

  }

\markboth{Journal of \LaTeX\ Class Files,~Vol.~18, No.~9, September~2020}%
{How to Use the IEEEtran \LaTeX \ Templates}

\maketitle

\begin{abstract}
Blind recognition of polar codes remains challenging in non-cooperative scenarios, particularly for information-set recognition with known code length. Existing methods mainly rely on threshold decisions determined by the generator-matrix structure and channel bit error probability, without fully exploiting the soft information in received signals. In this letter, 
we propose a blind recognition method using successive cancellation list (SCL) decoding for polar codes with known code length.
The proposed method exploits the distinct statistical behaviors of frozen and information bits in source-side decision log-likelihood ratios (LLRs) over multiple received vectors: frozen bits tend to favor zero decisions, whereas information bits exhibit nearly equiprobable $0/1$ decisions.
Based on this property, the decoder expands candidate paths under the frozen-bit and information-bit hypotheses at each bit position, evaluates their reliabilities using the corresponding average path metrics, and retains only the $L_{\mathrm{list}}$ most reliable paths for subsequent recognition. Finally, the information-set pattern corresponding to the most reliable surviving path is selected as the recognition result.
Simulation results show that the proposed scheme improves the recognition success rate as the list size increases. For the $(32,16)$, $(64,32)$, and $(128,64)$ polar codes, it achieves at least $2.5$ dB gain over the previous method when $L_{\mathrm{list}}=64$.
\end{abstract}


\section{Introduction}
\IEEEPARstart{P}{olar} codes, proposed by Ar{\i}kan, are the first class of channel codes proven to achieve the Shannon capacity of symmetric binary-input memoryless channels~[1]. Owing to their strong theoretical foundation and favorable performance--complexity tradeoff, polar codes have attracted considerable attention in modern communication systems. In particular, polar codes were adopted by 3GPP for 5G control-channel coding~[2], which has stimulated extensive research on their construction and decoding. Among various decoding algorithms, successive cancellation list (SCL) decoding has been widely studied because of its strong error-correction performance~[3], [4].

However, most existing studies on polar codes mainly focus on cooperative communication scenarios, where the receiver has full knowledge of the coding scheme and its parameters. In non-cooperative communication scenarios, by contrast, such prior information is usually unavailable, making blind recognition essential for subsequent decoding and signal analysis.

Existing studies on blind recognition of polar codes mainly follow two routes: hard decision and soft decision. In the hard-decision case,~[5] proposed a blind recognition method based on information-matrix estimation to identify the code length and code rate. In this method, decision thresholds are determined according to the generator-matrix structure and the channel error probability, so that frozen bits can be distinguished from information bits. This approach was further improved in~[6] through an estimation-and-derivation strategy, where the frozen-bit pattern identified at shorter code lengths is progressively propagated to longer candidate lengths for code-length and information-set identification. Since code-length recognition has become relatively reliable in existing studies,~[7] focused on the identification of information set and proposed a scheme combining multi-threshold voting with partial-order relations to improve reliability under high-BER conditions. In the soft-decision case,
[8] introduced log-likelihood ratios (LLRs) as soft information and exploited their statistical characteristics to detect the check relationships between the codewords and the suspected dual space over candidate code lengths, thereby identifying information set. This framework was further extended in~[9] by adopting the likelihood difference as the decision statistic, making it applicable to the recognition of polar codes with nonzero frozen bits.

Despite these advances, existing methods still have several limitations. Hard-decision-based schemes rely solely on binary decisions and ignore the reliability information contained in the channel output, resulting in limited performance under noisy conditions. Although soft-decision-based schemes exploit channel soft information, their recognition process still mainly depends on the structural properties of polar codes. As a result, the discriminative information carried by soft observations has not been fully utilized, leaving room for further improvement in recognition accuracy and robustness.

In this letter, we propose a blind successive cancellation list recognition (BSCL) scheme for polar codes with a known code length. The key innovation lies in seamlessly integrating channel soft information into the recursive bitwise path-search process inherent to SCL decoding. Specifically, at each bit position, candidate decoding paths are expanded under both the frozen-bit and information-bit hypotheses. 
To assess the reliability of each hypothesis, the corresponding path metrics are averaged over multiple received vectors. Based on these averaged metrics, only the $L_{\mathrm{list}}$ most reliable paths are retained for further recognition.
Finally, the information-set pattern that yields the minimum average path metric is selected as the recognition result. This approach enables robust blind recognition without requiring prior knowledge of the information set, effectively exploiting the inherent structure of polar codes.
\section{Background}

\subsection{System Model}
Let $\mathcal{P}(N,K)$ denote a polar code of length $N=2^n$ and dimension $K$. After channel polarization, the $K$ most reliable subchannels are assigned to information bits, while the remaining $N-K$ subchannels are assigned to frozen bits. Let $\mathcal{I}$ and $\mathcal{F}$ denote the information set and frozen set, respectively. For the source vector $\mathbf{u}=[u_0,u_1,\ldots,u_{N-1}]$, the corresponding codeword is generated as
\[
\mathbf{c}=\mathbf{u}\mathbf{G}_N,
\]
where $\mathbf{G}_N=\begin{bmatrix}1&0\\1&1\end{bmatrix}^{\otimes n}$ is the generator matrix.

Consider a non-cooperative scenario in which the receiver has no prior knowledge of the transmitter parameters, including the code length, code rate, and information set. Suppose that $M$ codewords, each of length $N$, are transmitted over an additive white Gaussian noise (AWGN) channel using binary phase-shift keying (BPSK) modulation. Then, the received symbol corresponding to the $j$-th coded bit of the $m$-th codeword is given by
\[
y_{m,j}=x_{m,j}+n_{m,j}, \quad 0\le m\le M-1,\; 0\le j\le N-1,
\]
where $x_{m,j}=1-2c_{m,j}\in\{+1,-1\}$ is the BPSK-modulated symbol corresponding to $c_{m,j}$, and $n_{m,j}\sim\mathcal{N}(0,\sigma^2)$ denotes the additive white Gaussian noise sample. The corresponding channel LLR is given by
\[
L(y_{m,j})=\frac{2y_{m,j}}{\sigma^2}.
\]
Collecting the channel LLRs of all $M$ received vectors yields
\[
\mathbf{L}=
\begin{bmatrix}
L(y_{0,0}) & L(y_{0,1}) & \cdots & L(y_{0,N-1})\\
L(y_{1,0}) & L(y_{1,1}) & \cdots & L(y_{1,N-1})\\
\vdots & \vdots & \ddots & \vdots\\
L(y_{M-1,0}) & L(y_{M-1,1}) & \cdots & L(y_{M-1,N-1})
\end{bmatrix},
\]
which serves as the input for subsequent blind recognition.

\subsection{SC/SCL Decoding of Polar Codes}

For a polar code of length $N$, SC decoding proceeds recursively over the decoding tree. 
For a length-$2$ polar code, the LLR associated with $u_0$ is first computed by the $f$-operation as
\begin{equation}
L(u_0)=f\big(L(y_0),L(y_1)\big)=L(y_0)\boxplus L(y_1),
\end{equation}
where
\begin{equation}
a\boxplus b = 2\operatorname{arctanh}\!\left(\tanh\frac{a}{2}\tanh\frac{b}{2}\right).
\end{equation}
After obtaining $\hat{u}_0$ by hard decision, the LLR associated with $u_1$ is computed through the $g$-operation as
\begin{equation}
L(u_1)=g\big(L(y_0),L(y_1),\hat{u}_0\big)=(-1)^{\hat{u}_0}L(y_0)+L(y_1).
\end{equation}
The hard decision rule is given by
\begin{equation}
\hat{u}_i=
\begin{cases}
0, & i\in\mathcal{F},\\
0, & i\in\mathcal{I},\ L(u_i)\ge 0,\\
1, & i\in\mathcal{I},\ L(u_i)<0.
\end{cases}
\end{equation}
By recursively applying the $f$- and $g$-operations over the decoding tree, SC decoding outputs the estimated source vector $\hat{\mathbf{u}}=[\hat{u}_0,\hat{u}_1,\ldots,\hat{u}_{N-1}]$.

SCL decoding follows the same recursive LLR update procedure as SC decoding, but preserves multiple candidate paths during the decoding process. At each frozen-bit position, only the path consistent with the frozen value is retained. At each information-bit position, each surviving path is expanded into two candidate paths corresponding to $\hat{u}_i=0$ and $\hat{u}_i=1$. The corresponding path metrics are updated, and only the $L_{\mathrm{list}}$ most reliable paths are retained for subsequent decoding.

\section{Blind Successive Cancellation List-Based Recognition Algorithm }

To address the unknown information set in non-cooperative scenarios, a blind SCL-based recognition decoder for polar codes is proposed. Unlike conventional polar SCL decoding, which expands decoding paths at information-bit positions, the proposed BSCL decoder expands paths over two hypotheses at each bit position, namely, the information-bit hypothesis and the frozen-bit hypothesis. Based on this idea, the principle of the BSCL decoder is first introduced, followed by its corresponding decoding procedure.

\subsection{Principle of the BSCL Algorithm }

In the considered non-cooperative scenario, all intercepted LLR streams are generated from codewords constructed by the same polar code and transmitted over the channel. Although the receiver does not know the information set, all intercepted observations correspond to codewords sharing the same information-bit pattern. Therefore, the resulting decision LLRs exhibit common statistical characteristics across different observations, which can be exploited for the blind recognition.

For the $m$-th intercepted observation, let $L(u_{m,j})$ denote the decision LLR of the $j$-th source bit, recursively computed from the $m$-th row of $\mathbf{L}$ through the $f$- and $g$-operations. In conventional SCL decoding, if the $j$-th source bit belongs to the frozen set, the corresponding path-metric increment is
\begin{equation}
\Lambda^{(\mathcal{F},j)}
=
\ln\!\left(1+e^{-L(u_{m,j})}\right),
\label{eq:metric_f_single}
\end{equation}
which measures the consistency between the decision LLR and the frozen-bit constraint. When $L(u_{m,j})\gg 0$, the decision strongly supports $u_{m,j}=0$, and the metric increment approaches zero. When $L(u_{m,j})=0$, the decision is neutral, and the increment equals $\ln 2$. In contrast, when $L(u_{m,j})\ll 0$, the decision tends to support $u_{m,j}=1$, so enforcing the frozen-bit constraint leads to a large metric penalty.
This property enables the conventional frozen-bit path metric to be exploited for blind  recognition. In general, if the current index does not belong to the information set, the resulting metric increment under the frozen-bit hypothesis tends to be smaller; otherwise, it tends to be larger.

Since all $M$ intercepted LLR observations are generated from codewords sharing the same information-bit pattern, identification based on a single observation may be sensitive to noise. To improve robustness, the frozen-bit hypothesis is evaluated jointly over all intercepted observations. Accordingly, the average metric increment under the frozen-bit hypothesis at index $j$ is defined as
\begin{equation}
\overline{\Lambda}^{(\mathcal{F},j)}
=
\frac{1}{M}\sum_{m=0}^{M-1}
\ln\!\left(1+e^{-L(u_{m,j})}\right),
\label{eq:metric_f_avg}
\end{equation}
which characterizes the overall consistency between the $j$-th index and the frozen-bit constraint over all intercepted observations. A smaller value of $\overline{\Lambda }^{(\mathcal{F},j)}$ indicates that the current index is more likely to be a frozen-bit position.

\textbf{Proposition 1:}
Under the Gaussian approximation (GA) assumption [10] and conditioned on correct preceding recursive decisions, let the decision LLR corresponding to a frozen-bit position with reliability parameter $\mu$ be modeled as
\[
X_\mu\sim\mathcal{N}(\mu,2\mu).
\]
Define
\begin{equation}
\Psi(\mu)=\mathbb{E}\!\left[\ln(1+e^{-X_\mu})\right],
\label{eq:Psi_def}
\end{equation}
which is the theoretical frozen-bit metric under the GA model. Then, $\Psi(\mu)$ is strictly decreasing with respect to $\mu$ for $\mu>0$.

\textit{Proof:}
Let
\[
g(x)=\ln(1+e^{-x}).
\]
Then
\[
g'(x)=-\frac{1}{1+e^x},\qquad
g''(x)=\frac{e^x}{(1+e^x)^2}.
\]
Hence,
\[
g'(x)+g''(x)
=
-\frac{1}{1+e^x}+\frac{e^x}{(1+e^x)^2}
=
-\frac{1}{(1+e^x)^2}<0.
\]
Since $\Psi(\mu)=\mathbb{E}[g(X_\mu)]$ with $X_\mu\sim\mathcal{N}(\mu,2\mu)$, using the equivalent representation
\[
X_\mu=\mu+\sqrt{2\mu}\,Z,\qquad Z\sim\mathcal{N}(0,1),
\]
and differentiating with respect to $\mu$, we obtain
\[
\Psi'(\mu)
=
\mathbb{E}[g'(X_\mu)]
+\frac{1}{\sqrt{2\mu}}\mathbb{E}[Zg'(X_\mu)].
\]
By Gaussian integration by parts,
\[
\frac{1}{\sqrt{2\mu}}\mathbb{E}[Zg'(X_\mu)]
=
\mathbb{E}[g''(X_\mu)].
\]
Therefore,
\[
\Psi'(\mu)=\mathbb{E}[g'(X_\mu)+g''(X_\mu)]<0.
\]
Thus, $\Psi(\mu)$ is strictly decreasing with respect to $\mu$. \hfill $\blacksquare$

\textbf{Corollary 1:}
Under the same condition as Proposition 1, for any frozen-bit position with reliability parameter $\mu>0$,
\[
\Psi(\mu)<\ln2.
\]

\textit{Proof:}
From Proposition 1, $\Psi(\mu)$ is strictly decreasing with respect to $\mu$. At $\mu=0$, we have
\[
X_0=0,
\]
and thus
\[
\Psi(0)=\ln(1+e^0)=\ln2.
\]
Therefore, for any $\mu>0$,
\[
\Psi(\mu)<\Psi(0)=\ln2.
\]
\hfill $\blacksquare$

Here, $\Psi(\mu)$ serves as a theoretical statistic for characterizing the frozen-bit metric under the GA assumption, while $\bar{\Lambda}^{(\mathcal{F},j)}$ denotes the empirical statistic used in the proposed algorithm. In particular, under favorable channel conditions, where the actual decoding behavior is closer to the GA assumption, sufficiently reliable frozen positions are more likely to satisfy $\bar{\Lambda}^{(\mathcal{F},j)} < \ln 2$.

Next, for the $m$-th intercepted observation, consider the path-metric increment under the information-bit hypothesis. In conventional SCL decoding, if the $j$-th source bit belongs to the information set, the current path is expanded into two candidate branches corresponding to the two possible bit decisions. The resulting path-metric increment can be written as
\begin{equation}
\Lambda ^{(\mathcal{I},j)}
=
\begin{cases}
\ln\!\left(1+e^{-L(u_{m,j})}\right), & \hat{u}_{m,j}=0,\\[4pt]
\ln\!\left(1+e^{L(u_{m,j})}\right), & \hat{u}_{m,j}=1,
\end{cases}
\label{eq:metric_info_nu}
\end{equation}
where $\hat{u}_{m,j}\in\{0,1\}$ denotes the candidate bit decision at position $j$. The candidate branch more consistent with the sign of $L(u_{m,j})$ tends to incur a smaller penalty. Therefore, if the conventional SCL information-bit metric is directly used for blind recognition, the information-bit hypothesis would be systematically favored. This is because the frozen-bit hypothesis always enforces $\hat{u}_{m,j}=0$, whereas the information-bit hypothesis can adapt to the instantaneous LLR and choose the more favorable branch. Hence, the conventional information-bit metric contains an inherent branch-selection advantage and cannot be directly used for a fair comparison between the frozen-bit and information-bit hypotheses.

For this reason, considering $M$ intercepted observations, a neutral reference metric is introduced for the information-bit hypothesis. According to Corollary~1, under the same GA assumption, the theoretical frozen-bit metric is strictly smaller than $\ln 2$ for any sufficiently reliable frozen position, while $\ln 2$ corresponds to the neutral case. Therefore, $\ln 2$ serves as a natural reference value for the information-bit hypothesis. Accordingly, the average information-bit metric increment at position $j$ is defined as
\begin{equation}
\overline{\Lambda}^{(\mathcal{I},j)}
=
\ln 2,
\label{eq:metric_i_single}
\end{equation}
which is used as the reference metric for the information-bit hypothesis in the subsequent blind-recognition procedure.

\subsection{Procedure of the BSCL Algorithm}
Based on the recognition principle in the previous subsection, a blind SCL-based recognition algorithm is proposed for information-set recognition from multiple intercepted observations. At each source index, every surviving path is expanded into two candidate branches under the frozen-bit hypothesis and the information-bit hypothesis, respectively, and only the best $L_{\mathrm{list}}$ paths are retained according to the resulting branch metrics. This list-search mechanism improves robustness when the channel Signal-to-Noise Ratio (SNR) is low or the number of intercepted observations is small, where single-path decisions may become unreliable and early recognition errors may propagate through the $f$- and $g$-operations.

For each surviving path, three types of path-dependent quantities are maintained throughout the recursion, namely, path labels, recursive LLR states, and partial sums. Here, the path labels record the hypothesized bit type at each processed source index, the recursive LLR states store the intermediate LLR messages required by the SCL recursion, and the partial sums store the temporary bit decisions required by the subsequent $g$-operations.

For the $m$-th intercepted observation, let $L(u_{m,j})$ denote the decision LLR at source index $j$, computed from its channel LLRs via the $f$- and $g$-operations. Let $\mathrm{PM}_l^{(j-1)}$ denote the accumulated metric of the $l$-th surviving path before processing source index $j$. Then, under the frozen-bit hypothesis,
\begin{equation}
\mathrm{PM}_l^{(j)}
=
\mathrm{PM}_l^{(j-1)}+\overline{\Lambda}_l^{(\mathcal{F},j)},
\label{eq:path_metric_f}
\end{equation}
and under the information-bit hypothesis,
\begin{equation}
\mathrm{PM}_l^{(j)}
=
\mathrm{PM}_l^{(j-1)}+\overline{\Lambda}_l^{(\mathcal{I},j)},
\label{eq:path_metric_i}
\end{equation}
where $\overline{\Lambda}_l^{(\mathcal{F},j)}$ and $\overline{\Lambda}_l^{(\mathcal{I},j)}$ are given by \eqref{eq:metric_f_avg} and \eqref{eq:metric_i_single}, respectively.

To continue the decoding recursion after path expansion, a temporary bit decision is assigned under each hypothesis. Under the frozen-bit hypothesis,
\begin{equation}
\hat{u}_{m,j}^{(\mathcal{F})}=0.
\label{eq:frozen_temp_decision}
\end{equation}
Under the information-bit hypothesis,
\begin{equation}
\hat{u}_{m,j}^{(\mathcal{I})}
=
\begin{cases}
0, & L(u_{m,j})\geq0,\\
1, & L(u_{m,j})<0.
\end{cases}
\label{eq:info_temp_decision}
\end{equation}
These temporary decisions are used only to update the partial sums required by the subsequent $g$-operations, and do not represent the final recognition output.

After all candidate branches at source index $j$ are generated, their accumulated metrics are sorted in ascending order, and only the best $L_{\mathrm{list}}$ branches are retained. The corresponding path labels, recursive LLR states, and partial sums are then updated for the retained branches. Repeating this procedure from $j=0$ to $j=N-1$ yields a set of candidate information-set patterns. Finally, the surviving path with the minimum accumulated metric is selected, and the estimated information set $\hat{\mathcal{I}}$ is recovered from its path-label sequence. The detailed procedure is summarized in Algorithm~\ref{alg:bscl_identification}.

Since the proposed BSCL scheme involves list-decoding operations, its computational complexity is higher than that of conventional blind identification methods. The overall complexity is \(\mathcal{O}(M L_{\mathrm{list}} N \log N)\), which mainly depends on the number of intercepted observations $M$ and the list size $L_{\mathrm{list}}$. Therefore, in practical applications, \(M\) and \(L_{\mathrm{list}}\) can be selected according to the desired tradeoff between identification performance and computational complexity.

\begin{algorithm}[t]
\caption{BSCL Algorithm}
\label{alg:bscl_identification}
\begin{algorithmic}[1]
\STATE \textbf{Input:} Channel LLR matrix $\mathbf{L}$ of $M$ intercepted observations, code length $N$, and list size $L_{\mathrm{list}}$
\STATE \textbf{Output:} Estimated information set $\hat{\mathcal{I}}$
\STATE Activate one initial path, set its accumulated metric to zero, and initialize the recursive LLR states, partial sums, and path labels
\FOR{$j=0$ to $N-1$}
    \FOR{each surviving path $l$}
        \FOR{$m=0$ to $M-1$}
            \STATE Compute the decision LLR $L(u_{m,j})$ via the $f$- and $g$-operations
            \STATE Set $\hat{u}_{m,j}^{(\mathcal{F})}$ according to \eqref{eq:frozen_temp_decision}
            \STATE Generate $\hat{u}_{m,j}^{(\mathcal{I})}$ according to \eqref{eq:info_temp_decision}
        \ENDFOR
        \STATE Evaluate $\overline{\Lambda}^{(\mathcal{F},j)}$ according to \eqref{eq:metric_f_avg}
        \STATE Set $\overline{\Lambda}^{(\mathcal{I},j)}$ according to \eqref{eq:metric_i_single}
        \STATE Generate candidate branches under the frozen-bit hypothesis and the information-bit hypothesis
    \ENDFOR
    \STATE Sort all candidate branches in ascending order of accumulated metric $\mathrm{PM}_l^{(j)}$
    \STATE Retain the best $L_{\mathrm{list}}$ branches
    \STATE Update the corresponding path labels, recursive LLR states, and partial sums
\ENDFOR
\STATE Select the surviving path with the minimum accumulated metric
\STATE Recover $\hat{\mathcal{I}}$ from its path-label sequence
\end{algorithmic}
\end{algorithm}

\section{Simulation Results}
In this section, the proposed blind recognition method is compared with the existing methods in [5], [6], [7], and [9]. In all simulations, the code length is assumed to be known, and all polar codes are constructed using the Gaussian approximation method at $E_b/N_0=2$ dB. The number of intercepted observations is denoted by $M$, and the number of Monte Carlo trials is set to 1000. The symbol $P_l$ denotes the recognition success rate. The base codewords are BPSK-modulated and transmitted over an AWGN channel. In Figs.~1--3, the SNR is represented by $E_s/N_0$, and the corresponding channel bit error probability is $P_b=Q\!\left(\sqrt{2E_s/N_0}\right)=Q\!\left(\sqrt{2\cdot10^{\frac{(E_s/N_0)_{\mathrm{dB}}}{10}}}\right)$.

\begin{figure}[t]
  \centering
  \includegraphics[width=1\linewidth]{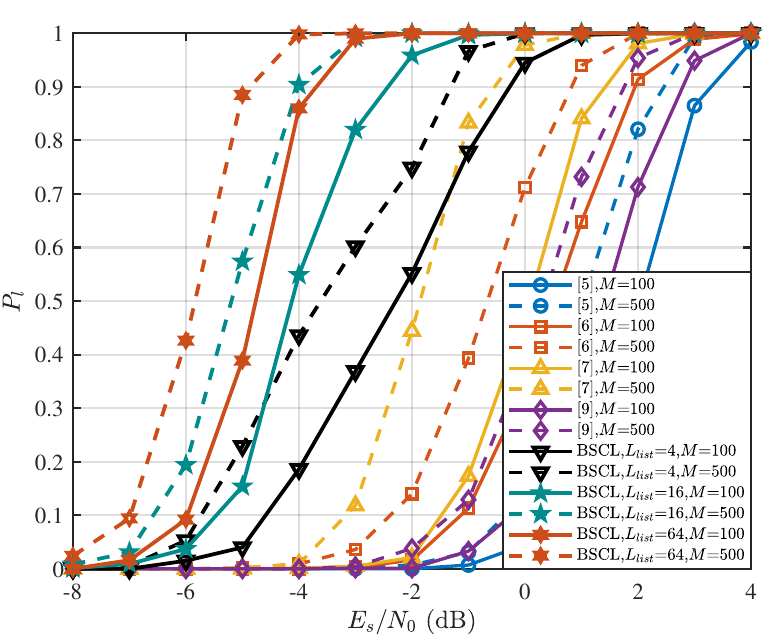}
  \caption{ Performance comparison with $\mathcal{P}(32,16)$ }
  \label{fig:example}
\end{figure}

\begin{figure}[t]
  \centering
  \includegraphics[width=1\linewidth]{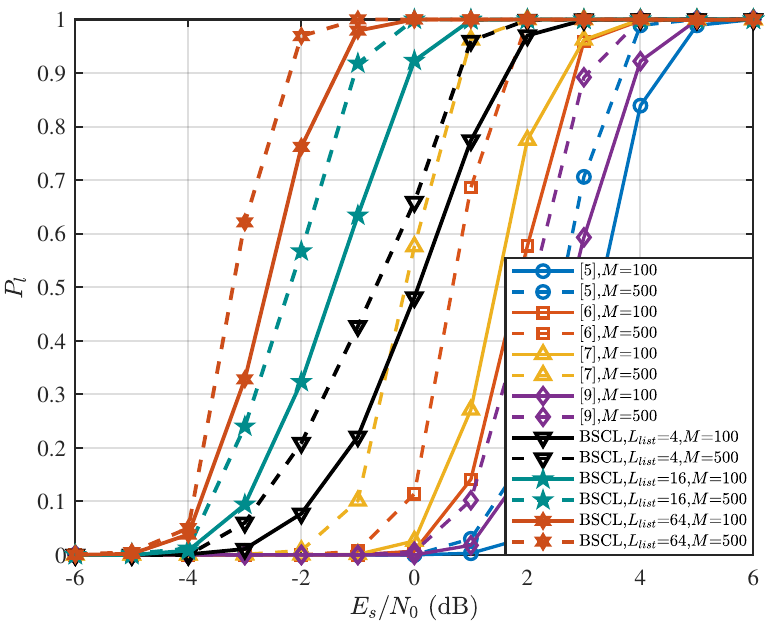}
  \caption{Performance comparison  with $\mathcal{P}(64,32)$}
  \label{fig:example}
\end{figure}

\begin{figure}[t]
  \centering
  \includegraphics[width=1\linewidth]{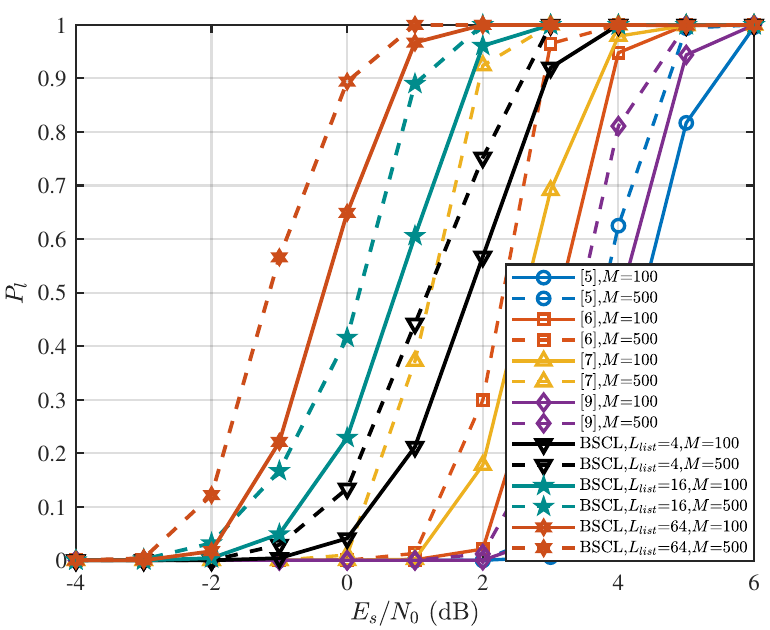}
  \caption{Performance comparison with $\mathcal{P}(128,64)$ }
  \label{fig:example}
\end{figure}

Fig.~1 compares the recognition success rates of the $\mathcal{P}(32,16)$ polar code for different schemes. As the list size increases, the proposed BSCL scheme achieves improved recognition performance. For $L_{\mathrm{list}}=64$, the proposed BSCL scheme shows a clear advantage over the best existing scheme in [7]. Specifically, when the number of intercepted codewords is $M=100$, it achieves about $5$ dB gain at $P_l=0.6$. When $M=500$, it still provides about $4.0$ dB gain at $P_l=0.8$.
Fig.~2 compares the recognition success rates of the $\mathcal{P}(64,32)$ polar code for different schemes. As the list size increases, the proposed BSCL scheme achieves improved recognition performance. For $L_{\mathrm{list}}=64$, the proposed BSCL scheme shows a clear advantage over the best existing scheme in [7]. Specifically, when the number of intercepted codewords is $M=100$, it achieves about $4$ dB gain at $P_l=0.7$. When $M=500$, it still provides about $3.0$ dB gain at $P_l=0.9$.
Fig.~3 compares the recognition success rates of the $\mathcal{P}(128,64)$ polar code for different schemes. As the list size increases, the proposed BSCL scheme achieves improved recognition performance. For $L_{\mathrm{list}}=64$, the proposed BSCL scheme shows a clear advantage over the best existing scheme in [7]. Specifically, when the number of intercepted codewords is $M=100$, it achieves about $3$ dB gain at $P_l=0.8$. When $M=500$, it still provides about $2.5$ dB gain at $P_l=0.5$.

\section{Conclusion}
In this letter, a blind SCL-based recognition scheme for polar codes with known code length was proposed. By introducing frozen-bit and information-bit hypotheses as competing decoding paths, the proposed scheme establishes an effective connection between polar decoding and blind recognition, while making full use of the received soft information. Simulation results demonstrated that the proposed scheme significantly improves the recognition success rate, and achieves clear gains over the existing method as the list size increases.


\begin{thebibliography}{1}


\bibitem{b1}
E. Ar{\i}kan, “Channel polarization: A method for constructing capacity-achieving codes for symmetric binary-input memoryless channels,” \emph{IEEE Trans. Inf. Theory}, vol. 55, no. 7, pp. 3051--3073, Jul. 2009.

\bibitem{b2}
3rd Generation Partnership Project (3GPP), “TSG RAN WG1 AH1 NR,” Athens, Greece, Feb. 2017.

\bibitem{b3}
I. Tal and A. Vardy, “List decoding of polar codes,” \emph{IEEE Trans. Inf. Theory}, vol. 61, no. 5, pp. 2213--2226, May 2015.

\bibitem{b4}
K. Niu and K. Chen, “CRC-aided decoding of polar codes,” \emph{IEEE Commun. Lett.}, vol. 16, no. 10, pp. 1668–1671, Oct. 2012.

\bibitem{b5}
J. Liu, T. Zhang, H. Bai, and S. Ye, “Blind recognition algorithm of polar code based on information matrix estimation,” \emph{Syst. Eng. Electron.}, vol. 44, no. 2, pp. 668--676, 2022.

\bibitem{b6}
C. Yi, B. Pang, L. He, B. Ma, Y. Li, and F. C. M. Lau, “Blind identification of polar codes based on estimation and derivation approaches,” \emph{IEEE Commun. Lett.}, vol. 27, no. 2, pp. 414--418, Feb. 2023.

\bibitem{b7}
P. Xu, J. Liu, A. Wang, C. Yi, and Q. Li, “Blind recognition of polar code information bits based on multi-threshold voting and partial orders,” \emph{IEEE Commun. Lett.}, vol. 30, pp. 887--891, 2026.

\bibitem{b8}
Z. Wu, Z. Zhong, L. Zhang, and B. Dan, “Recognition of non-drilled polar codes based on soft decision,” \emph{J. Commun.}, vol. 41, no. 12, pp. 60--71, Dec. 2020.

\bibitem{b9}
Y. Wang, C. Wang, X. Wang, and Z. Huang, “Non-punctured polar code parameter recognition algorithm based on soft decision,” \emph{Syst. Eng. Electron.}, vol. 45, no. 10, pp. 3293--3301, Oct. 2023.

\bibitem{b10}	
P. Trifonov, “Efficient design and decoding of polar codes,” \emph{IEEE Trans. Commun.}, vol. 60, no. 11, pp. 3221–3227, Nov. 2012.

\end{thebibliography}
\end{document}